\documentclass{PoS}

\title{The Microchannel X-ray Telescope on Board the SVOM Satellite}

\ShortTitle{The Microchannel X-ray Telescope}

\author{\speaker{D. G\"otz}$^a$, C. Adami$^b$, S. Basa$^b$, V. Beckmann$^c$, V. Burwitz$^d$, R. Chipaux$^e$, B. Cordier$^a$, P. Evans$^f$, O. Godet$^g$, R. Goosmann$^h$, N. Meidinger$^d$, A. Meuris$^a$, C. Motch$^h$, K. Nandra$^d$, P. O'Brien$^f$, J. Osborne$^f$, E. Perinati$^i$, A. Rau$^d$, R. Willingale$^f$, K. Mercier$^j$, F. Gonzalez$^j$\\
        \llap{$^a$}CEA Saclay -- DSM/Irfu/Service d'Astrophysique, F-91191, Gif-sur-Yvette, France\\
        \llap{$^b$}Aix Marseille Universit\'e, CNRS, LAM (Laboratoire d'Astrophysique de Marseille) UMR 7326, 13388, Marseille, France\\
        \llap{$^c$}Fran\c{c}ois Arago Centre, APC, Universit\'e Paris Diderot, CNRS/IN2P3, CEA/Irfu, Observatoire de Paris, Sorbonne Paris Cit\'e, 13 Rue Watt, 75013 Paris, France\\
        \llap{$^d$}MPE, Giessenbachstrasse, D-85748 Garching, Germany\\
        \llap{$^e$}CEA Saclay -- DSM/Irfu/SEDI, F-91191, Gif-sur-Yvette, France\\
        \llap{$^f$}Dept. of Physics \& Astronomy -- University of Leicester, Leicester LE1 7RH, GB\\
        \llap{$^g$}Institut de Recherche en Astrophysique and Plan\'etologie (IRAP), Universit\'e de Toulouse, UPS, 9 Avenue du colonel Roche, F-31028 Toulouse Cedex 4, France\\
        \llap{$^h$}Observatoire astronomique de Strasbourg, Universit\'e de Strasbourg, CNRS, UMR 7550, 11
rue de l'Universit\'e, F-67000 Strasbourg, France\\
        \llap{$^i$}Institut f\"ur Astronomie und Astrophysik T\"ubingen, Eberhard Karls Univ. T\"ubingen, Germany\\
        \llap{$^j$}CNES, 18 avenue Edouard Belin, F-31400 Toulouse, France

        E-mail: \email{diego.gotz@cea.fr}}

\abstract{We present the Micro-channel X-ray Telescope (MXT), a new narrow-field (about 1$^\circ$) telescope that will be flying on the Sino-French SVOM mission dedicated to Gamma-Ray Burst science, scheduled for launch in 2021. MXT is based on square micro pore optics (MPOs), coupled with a low noise CCD. The optics are based on a ``Lobster Eye'' design, while the CCD is a focal plane detector similar to the type developed for the seven eROSITA telescopes. MXT is a compact and light ($<$35 kg) telescope with a 1 m focal length, and it will provide an effective area of about 45 cm$^2$ on axis at 1 keV. The MXT PSF is expected to be better than 4.2 arc min (FWHM) ensuring a localization accuracy of the afterglows of the SVOM GRBs to better than 1 arc min (90\% c.l. with no systematics) provided MXT data are collected within 5 minutes after the trigger. The MXT sensitivity will be adequate to detect the afterglows for almost all the SVOM GRBs as well as to perform observations of non-GRB astrophysical objects. These performances are fully adapted to the SVOM science goals, and prove that small and light telescopes can be used for future small X-ray missions.}

\FullConference{Swift: 10 Years of Discovery,\\
		2-5 December 2014\\
		La Sapienza University, Rome, Italy }

\begin{document}

\section{Introduction}
SVOM (Space based astronomical Variable Object Monitor) is a Sino-French mission dedicated to Gamma-Ray Bursts (GRBs) studies developed in cooperation by the French Space Agency (CNES), the Chinese National Space Administration (CNSA) and the Chinese Academy of Sciences (CAS). SVOM will be launched in 2021, and will carry two wide field instruments -- the coded mask telescope ECLAIRs (4--150 keV, see \cite{godet} and Schanne et al. these proceedings), providing the GRB triggers, and a non-imaging gamma-ray spectrometer GRM (50 keV--5 MeV), and two narrow field instruments -- a visible telescope, VT, and the MXT X-ray telescope, that are pointed at the ECLAIRs error box after an autonomous platform slew to observe the GRB afterglows. SVOM alerts will be transmitted to ground through a VHF network and will be delivered to the scientific community within 30 seconds from detection in most of the cases. SVOM will also include a set of ground based instruments: the GWACs, an ensemble of wide angle cameras that will continuously monitor the central part of the field of view of ECLAIRs, in order to catch the GRB prompt emission in the visible band, and two dedicated robotic ground follow-up telescopes (GFTs). The latter will be provided by the two partners and installed in China and Mexico.
For more information about the SVOM mission, see \cite{paul} and Cordier et al. these proceedings.

\section{The Telescope and the Optics}
\begin{figure}[ht!]
\center\includegraphics[width=.45\textwidth]{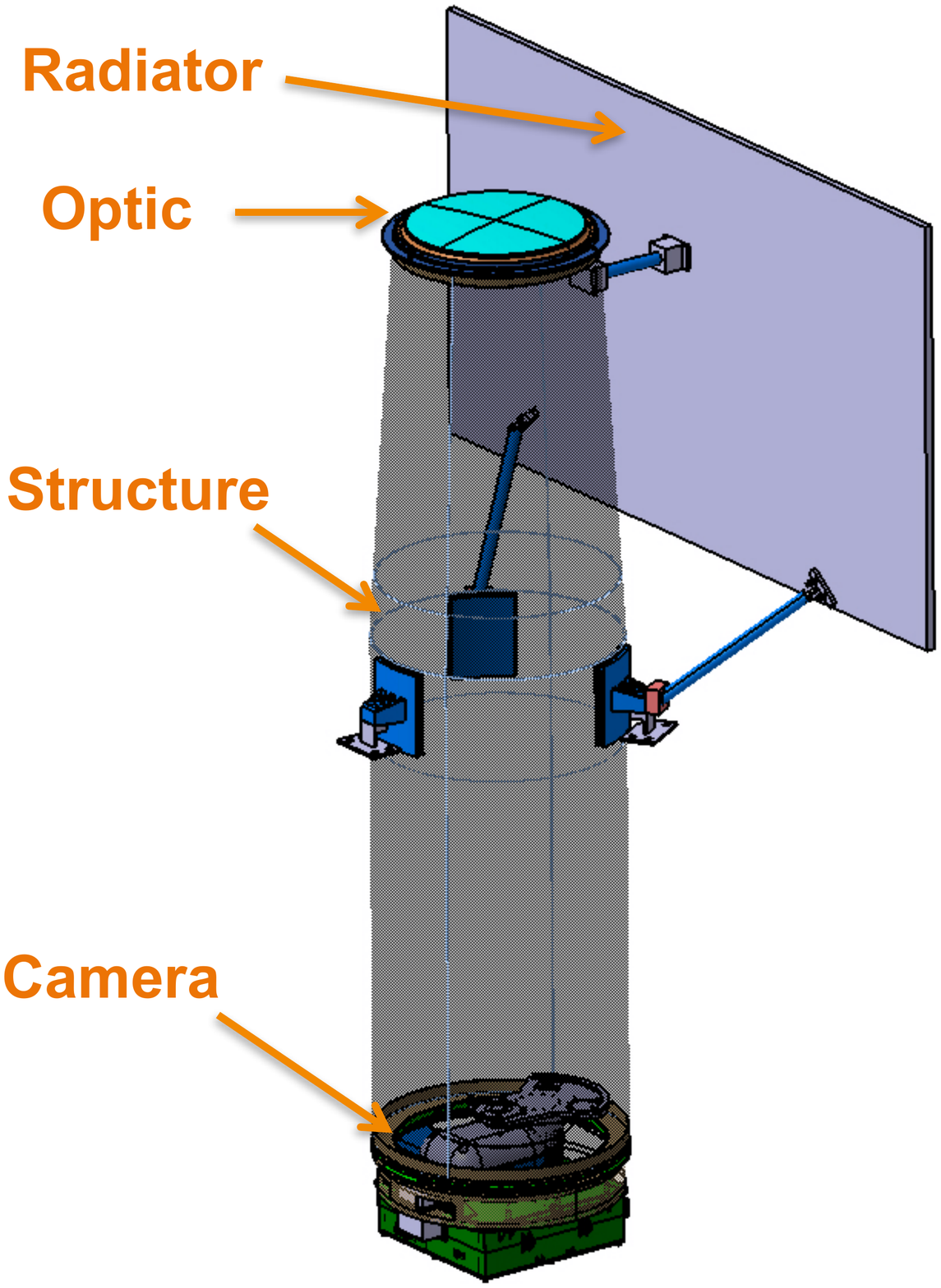}
\caption{The MXT telescope and its subsystems.}
\label{fig:mxt}
\end{figure}

The Microchannel X-ray Telescope (MXT) is developed under CNES responsibility in close collaboration with CEA/Irfu, the University of Leicester, and the Max-Planck-Institut f\"ur Extraterrestrische Physik (MPE). It will be composed of the main subsystems indicated in Fig. \ref{fig:mxt}, namely an optic based on square micro-channels, a camera, a carbon fibre structure, and a radiator.

The optical scheme of the MXT is based on a ``Lobster Eye'' grazing incidence geometry as defined by \cite{angel}, and optimised for narrow-field use. Reflections occur within the square pores of square micro-pore glass plates (MPOs), which have a thickness of 1-2 mm and a side length of 40 mm; these plates, produced by Photonis, are spherically slumped to a radius twice the focal length of the MXT.

\begin{figure}[ht!]
\includegraphics[width=.45\textwidth]{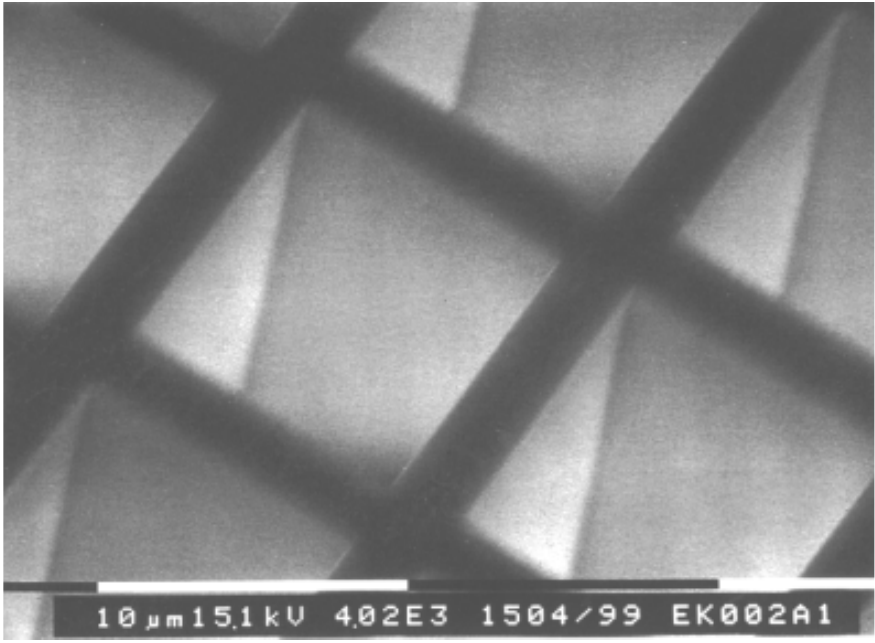}
\hspace{1.5cm}\includegraphics[width=.45\textwidth]{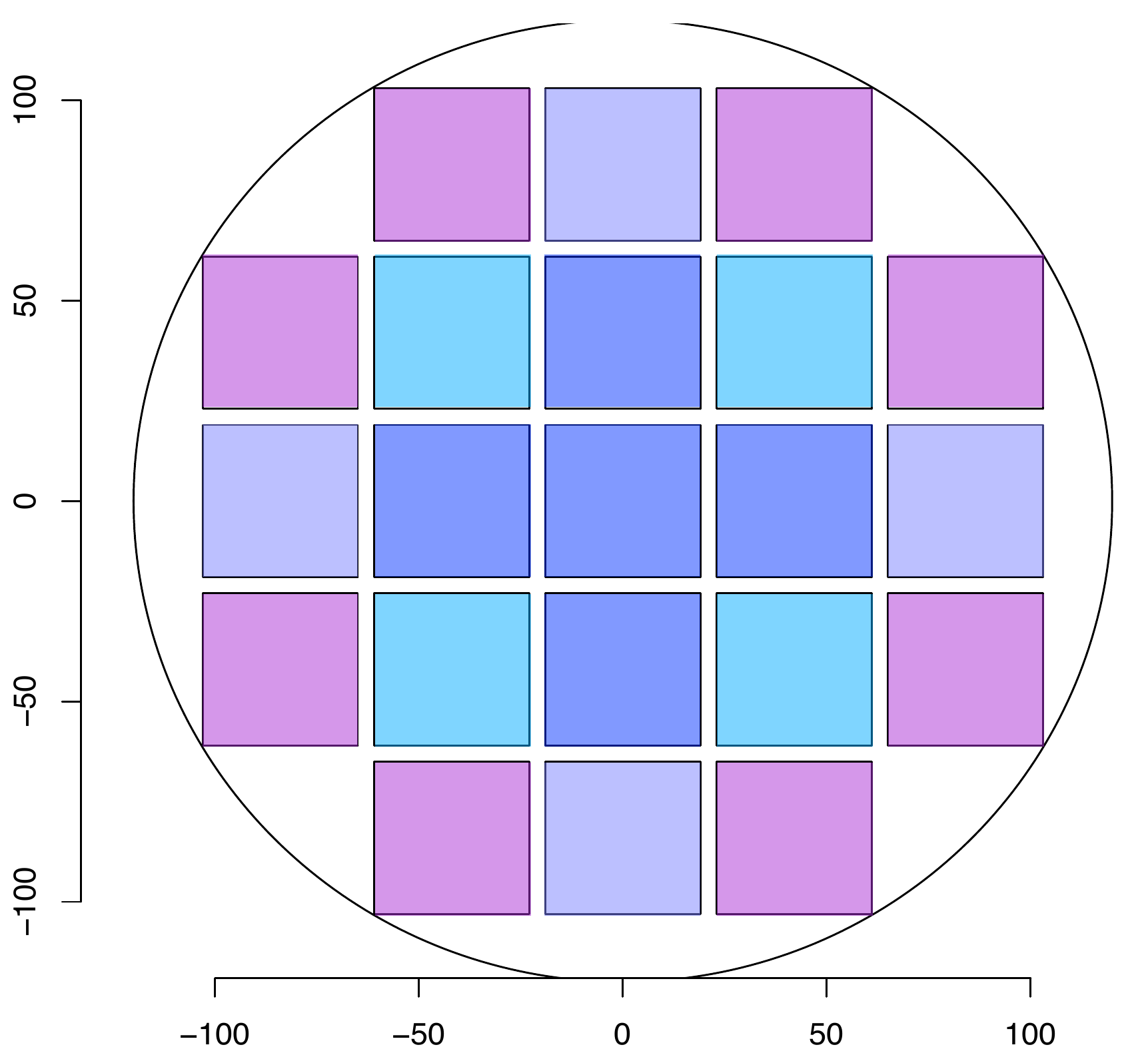}
\caption{Left: Close-up view of a square glass MPO of 20 $\mu$m side, from \cite{fraser}. MPOs are coated with a thin (few tens of $\mu$m) layer of a high Z element (e.g. Platinum, Iridium, etc.) to boost their X-ray reflectivity, especially at higher energies. Right: The square MPOs of different length (1-2 mm, colour coded) are mounted on a frame of 24 cm of diameter (axes are in mm): 21 MPOs are needed to form the MXT optic.}
\label{fig:opt2}
\end{figure}

\begin{figure}[ht!]
\includegraphics[width=.45\textwidth]{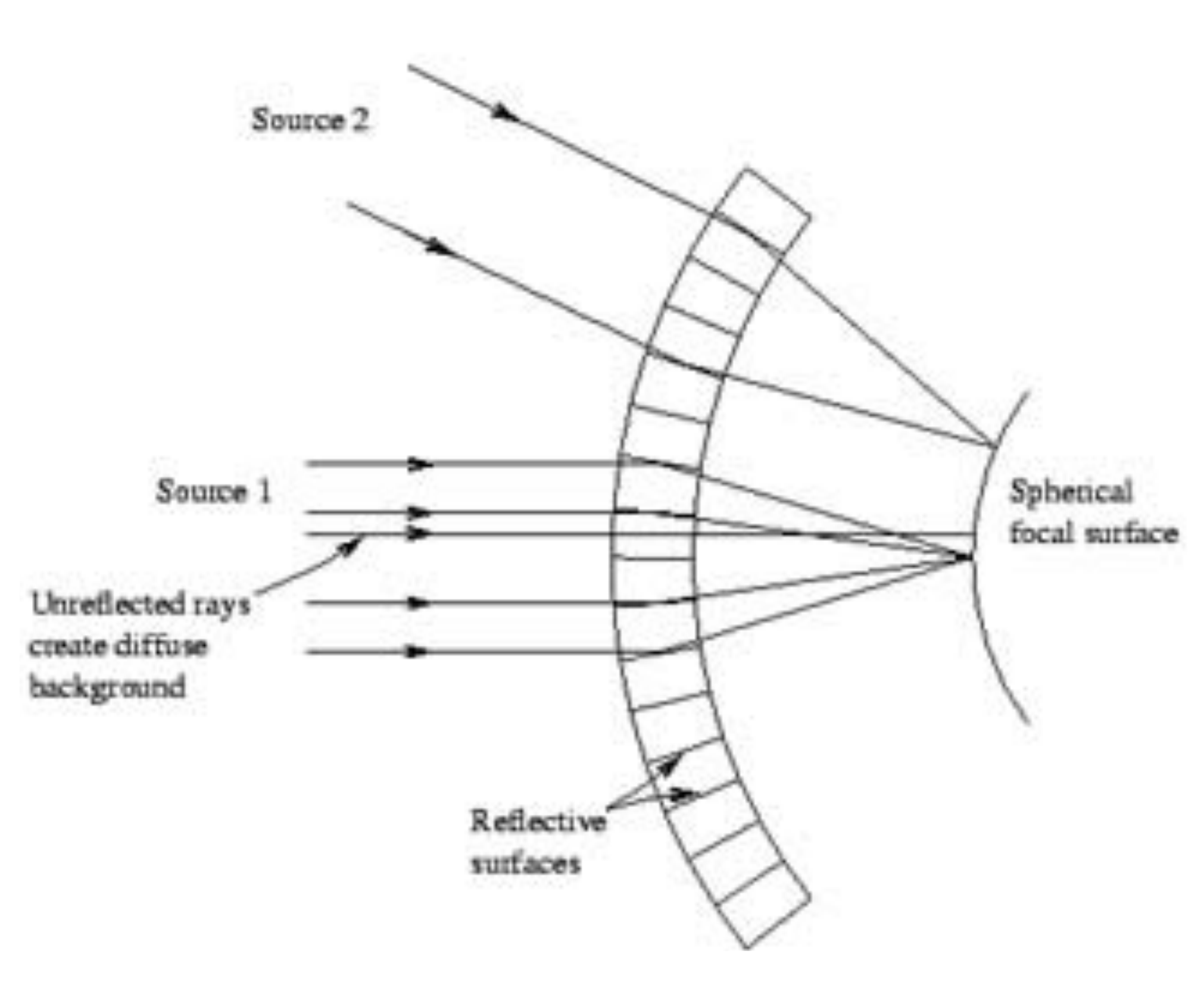}
\hspace{1cm}\includegraphics[width=.45\textwidth]{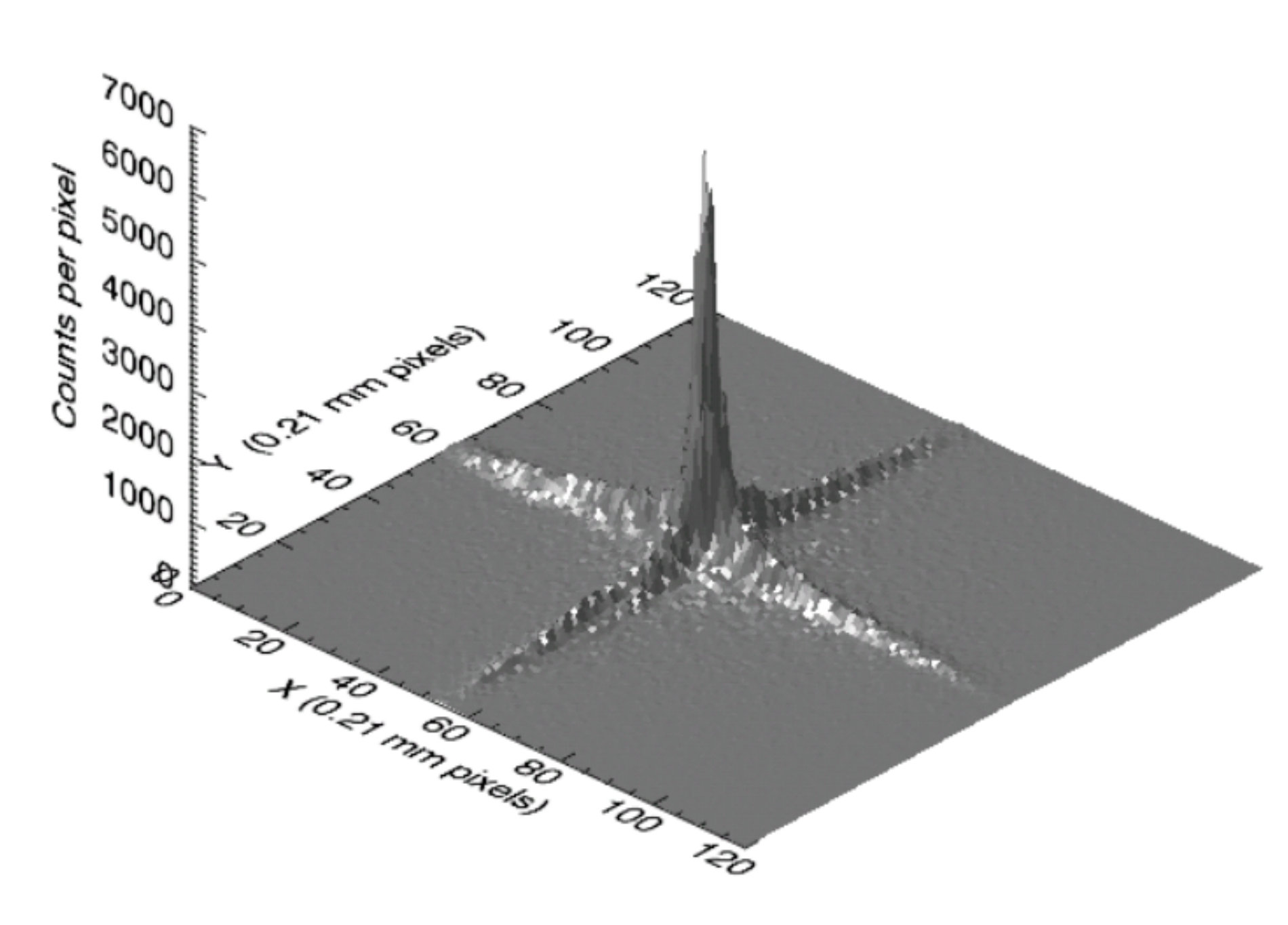}
\caption{Left: Working principle of a ``Lobster Eye" X-ray optic. Right: Simulated MXT point spread function with the characteristic central peak, cross arms and diffuse patch, see text.}
\label{fig:opt1}
\end{figure}

The MPOs will be coated with Ir and mounted on a round frame, see Fig. \ref{fig:opt2}, in order to obtain a complete X-ray optic with a 1 m focal length and coverage of the 64$\times$64 arc min field of view defined by the detector. The front of the optic will be coated with 70 nm of Al for thermal reasons and in order to reduce the optical load on the detector.

The ``Lobster Eye'' optic was originally developed in order to cover very large areas of the sky with homogeneous sensitivity. Thanks to its light weight with respect to classical X-ray optics, it is very well adapted to mini class satellites like SVOM. Its working principle is sketched in Fig. \ref{fig:opt1}. X-rays entering in the MPOs can either be reflected twice and focused in the central PSF spot, or reflected just once and focused in the PSF arms. 25\% of the incident X-ray flux is focussed in the central spot, 2$\times$25\% in the arms, and the rest in a diffuse patch, see Fig. \ref{fig:opt1}. Thanks to the ``Lobster Eye'' geometry the vignetting is very low, of the order of 15\% at the edge of the FOV. Current simulations indicate that the PSF FWHM of such a system is 4.2 arc min at 1 keV.

\section{MXT Camera}

\begin{figure}[ht!]
\includegraphics[width=.45\textwidth]{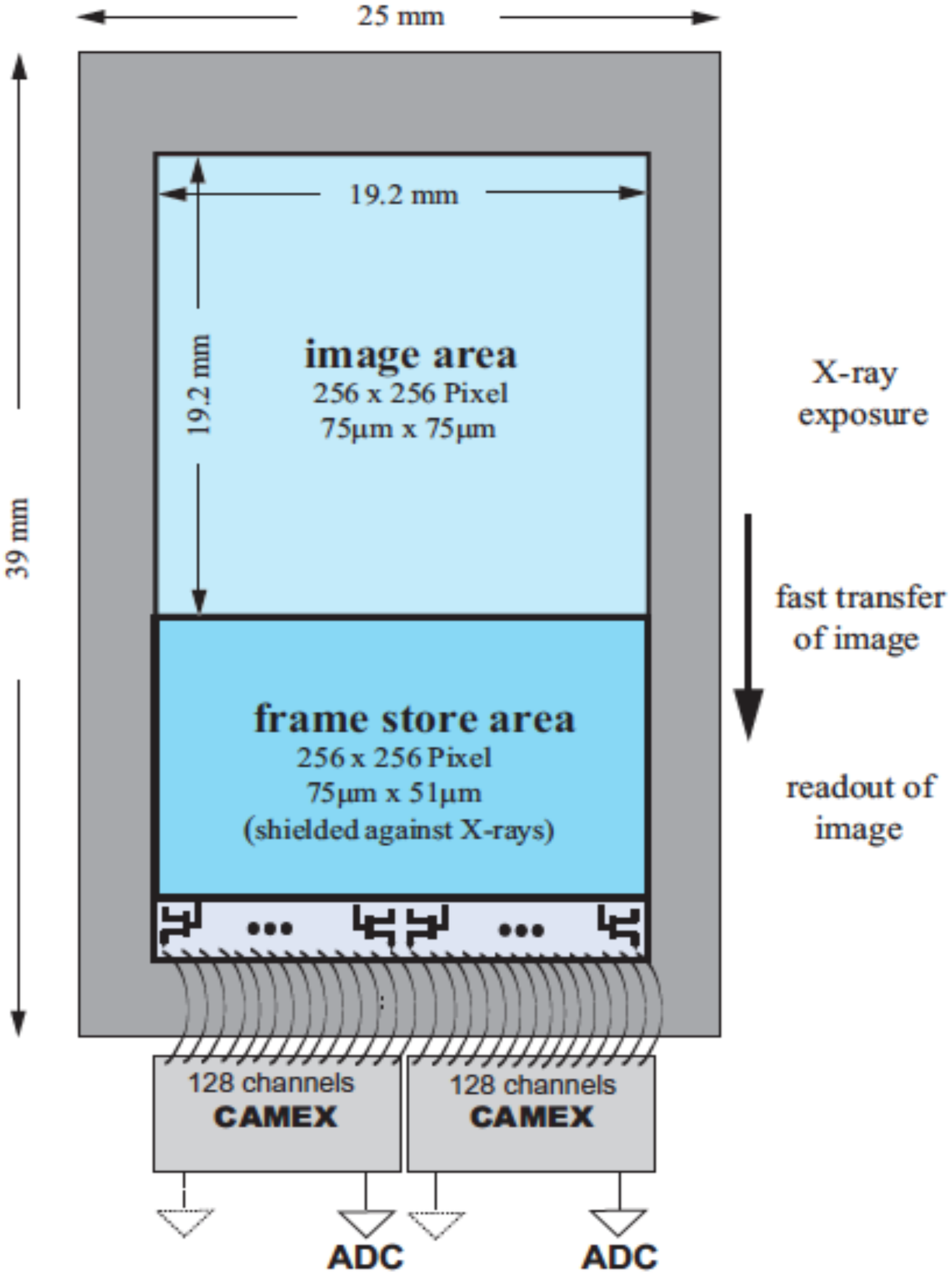}
\hspace{1.5cm}\includegraphics[width=.45\textwidth]{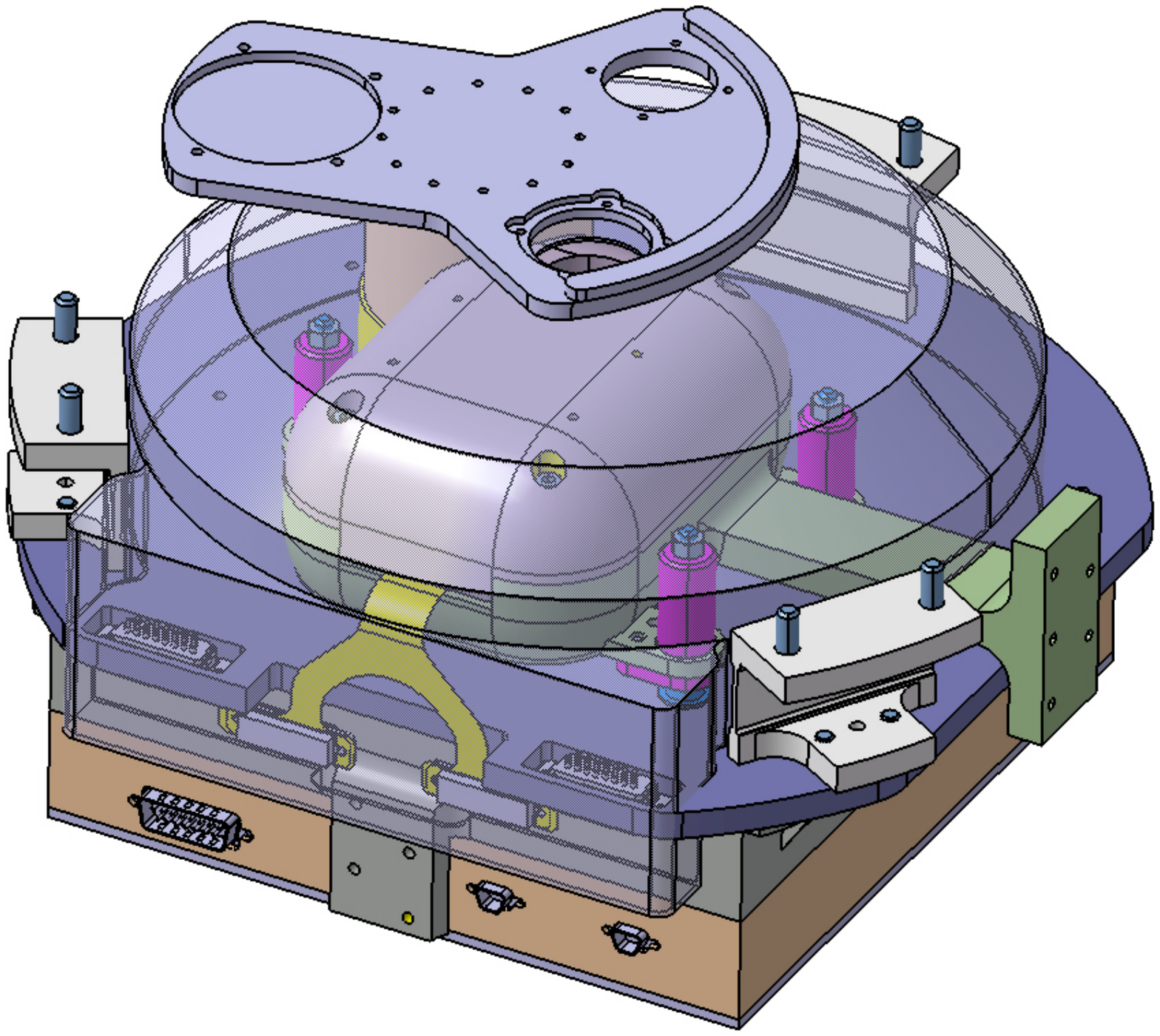}
\caption{Left: Conceptual scheme of the pnCCD that will be used for MXT. Right: Mechanical structure of the MXT camera.}
\label{fig:mcam}
\end{figure}

The MXT camera design is based on a pnCCD developed by the MPE/HLL \cite{meidinger}, and is a small scale version of the detectors that will be used in the eROSITA \cite{predehl} instrument on board the Spectrum R\"ontgen-Gamma satellite, see Fig. \ref{fig:mcam}. The pnCCD has an active area of 256$\times$256 pixels of 75 $\mu$m, and a reduced frame store area with 75$\times$51 $\mu$m pixel. The CCD is fully depleted (450 $\mu$m depth) and has excellent low-energy response (45-48 eV (FWHM) @ 277 eV), and energy resolution (123-131 eV FWHM @ 5.9 keV). It is read out by two CAMEX ASICs  providing a very low read-out noise (2.18 e$^-$ ENC @ -- 60$^\circ$C). 
The CAMEX ASIC permits a parallel and thus fast readout of all channels of the frame transfer pnCCD.
The detector will be shielded by 3 cm of equivalent Al, and cooled by the thermo electric cooler (TEC) to -65$^\circ$ C. The camera will include digital and analogue front-end electronics, to drive the CCD and to perform its readout, and a filter wheel to be used as a shutter to protect the CCD during SAA passages, and to put a calibration source or and additional optical/UV filter in front of the detector when needed. The GRB afterglow position will be computed in real-time, on board, by the MXT Data Processing Unit (M-DPU).

\section{Scientific Performance}
\begin{figure}[ht!]
\includegraphics[width=.45\textwidth]{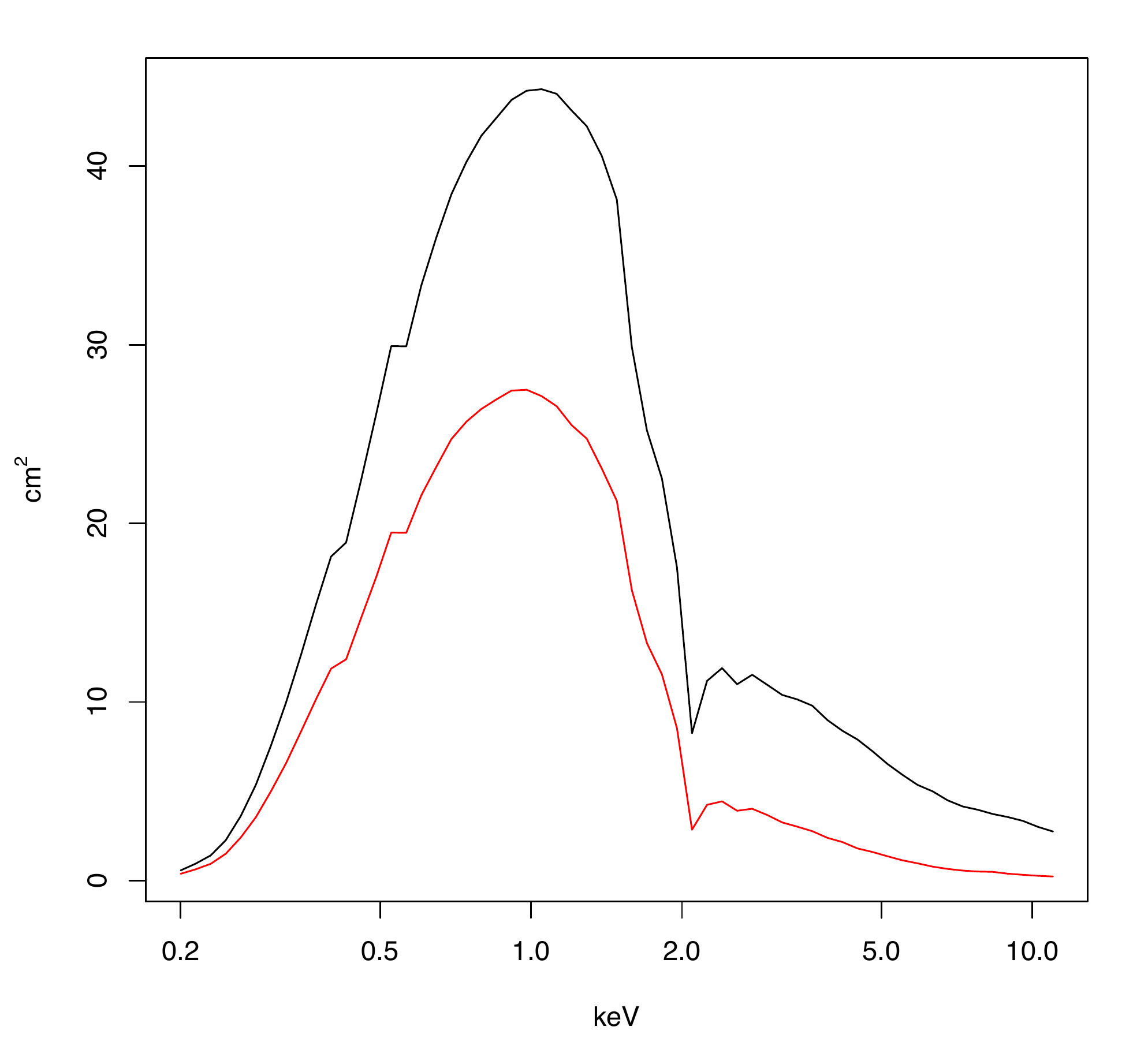}
\includegraphics[width=.55\textwidth]{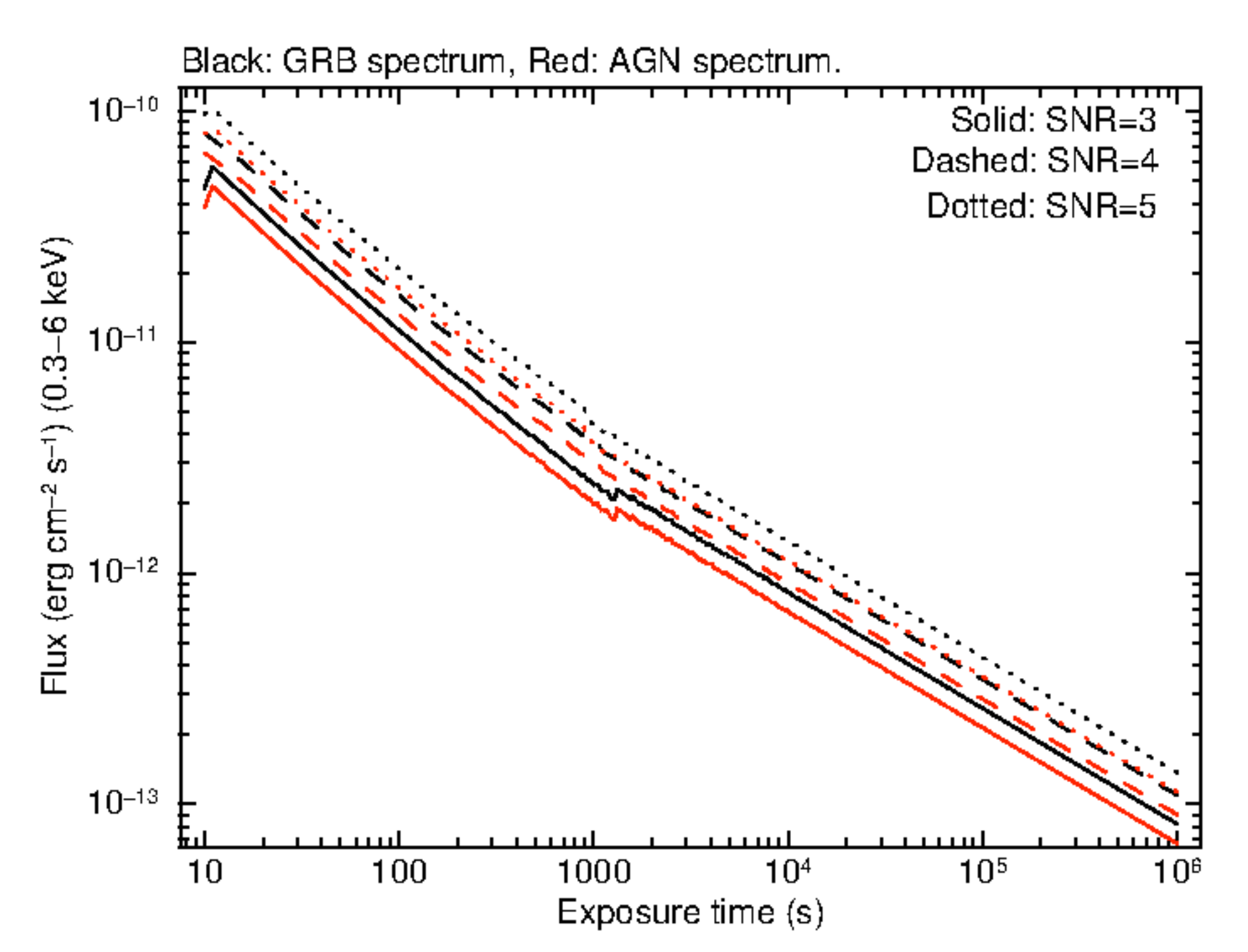}
\caption{Left: MXT on-axis central spot (red) and spot+arms+patch (black) effective area for a 1 m focal length telescope. For a 1.15 m, the effective area is expected to increase by about 20\%.  The MXT area curve is derived from the pnCCD quantum efficiency, and an assumed 200 nm Al filter. Right: Estimated MXT sensitivity as a function of exposure time. The curves show the 0.3-6 keV absorbed flux detection limits for 3, 4 and 5 sigma significance above background as a function of integration time for a detection and background cell size fixed at 80\% of the enclosed PSF. The black curve is obtained for a photon index $\Gamma$=2.0, and an absorption column density $N_{H}$=3$\times$10$^{21}$ atoms cm$^{-2}$, while the red one for $\Gamma$=1.7 and $N_{H}$=3$\times$10$^{20}$ atoms cm$^{-2}$.}
\label{fig:perf1}
\end{figure}

\begin{figure}[ht!]
\includegraphics[width=.5\textwidth]{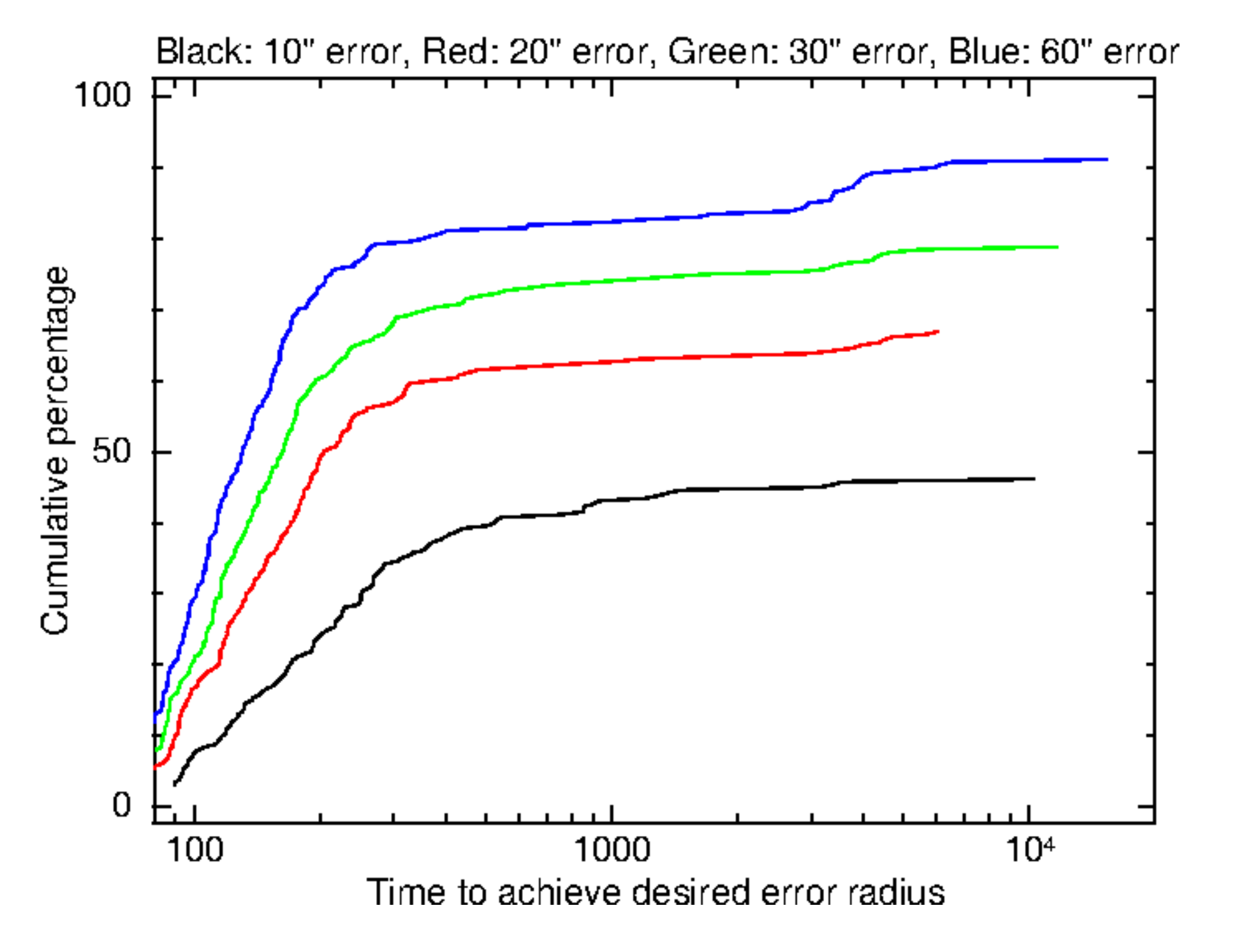}
\includegraphics[width=.5\textwidth]{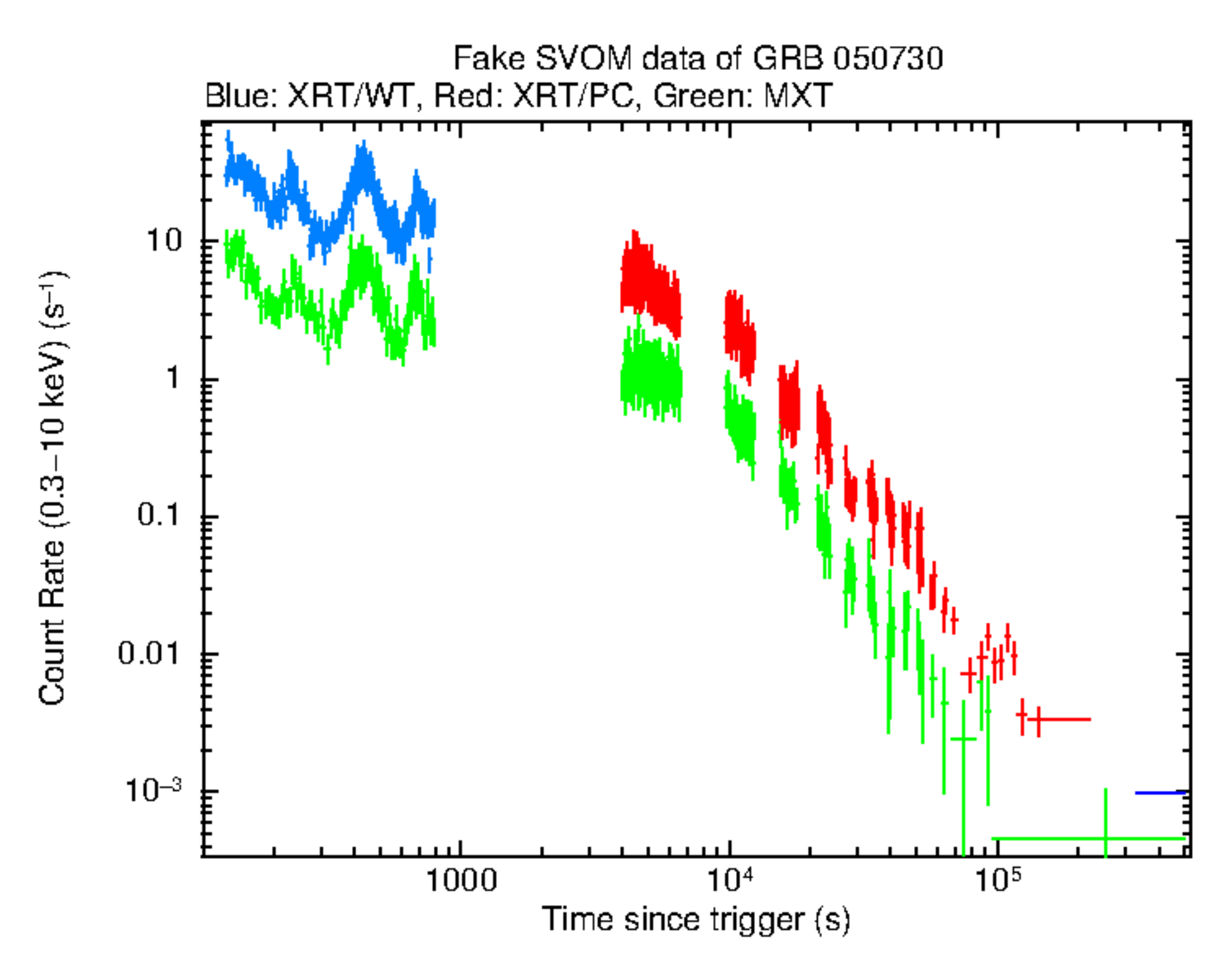}
\caption{Left: The cumulative distributions of the various 90\% confidence level error radii as a function of time since trigger for all bursts to which SVOM has slewed. Right: MXT simulated (green) vs. Swift XRT (blue and red) light curve of a GRB afterglow with a median flux over the first orbit in the XRT archive.}
\label{fig:perf2}
\end{figure}

The MXT expected performances are illustrated in the Figures \ref{fig:perf1} and \ref{fig:perf2}. 
The MXT expected effective area at 1 keV is 27 cm$^2$ for the central spot and 44 cm$^2$ if one includes also the PSF cross arms. If one takes these values into account, and includes the Cosmic X-ray Background as well as the expected particle background, one obtains a 5$\sigma$ sensitivity for MXT in the 0.3--6 keV energy band of of 8$\times$10$^{-11}$ erg cm$^{-2}$ s$^{-1}$ for a 10 s observation, and $\sim$10$^{-12}$ erg cm$^{-2}$ s$^{-1}$ for a 10 ks one.
 
Using the values obtained above, we estimated the MXT scientific performance with respect to GRB afterglows observations by folding the the entire Swift/XRT afterglow data set through the MXT response (assuming a 4.2 arc min PSF at 1 keV). One can see that the MXT is fully adapted to the study of SVOM afterglows. Indeed 50\% of the bursts will be located to better than 13 arc sec (statistical uncertainties only) within 5 minutes from trigger. In addition, despite the smaller effective area with respect to XRT, most of the afterglows will be detected up to 10$^5$ seconds after trigger, with a good signal-to-noise ratio, see Fig. \ref{fig:perf2}.

We also investigated the possibility to increase the focal length of the telescope from 1 to 1.15 m. This is the maximum value which is allocated to MXT on the SVOM satellite. Preliminary results indicate that the central spot effective area will increase to about 32 cm$^{2}$, and that the sensitivity of the telescope may increase by 10--20\% as a function of the source column density (N$_{H}$). This increased effective area implies for afterglow observations that the same localisation accuracy will be reached in about 20\% less of the observing time. The impact of the increased focal length is the reduction of the MXT field of view to 58$\times$58 arc min (overall reduction of 24\% of the projected sky area), which is still large enough to include the largest error boxes provided by ECLAIRs. The possible technical implementation of this solution requires some update of the telescope design, and will be evaluated during phase B.

\section{Conclusions}
MXT is a very light ($\sim$35 kg), and compact ($\sim$1.2 m) focussing X-ray telescope. Its large field of view ($\sim$1 degree) and its sensitivity below the mCrab level make of MXT a very good instrument to identify and precisely localize (below the arc minute) X-ray transients in non-crowded fields, and to study them in detail, thanks to its excellent spectral response.

\end{document}